\documentclass[aps,pra,showpacs,twocolumn]{revtex4} 
\usepackage{epsfig}  
\draft 
\begin{document} 
 
\title{Distinguishing mixed quantum states: Minimum-error discrimination  versus\\  
optimum unambiguous discrimination}  
\author{Ulrike Herzog$^1$} 
\author{J\'{a}nos A. Bergou$^2$} 
\affiliation{$^1$Institut f\"ur Physik,  Humboldt-Universit\"at  
  zu Berlin, Newtonstrasse 15, D-12489 Berlin, Germany} 
\affiliation{$^2$Department of Physics and Astronomy,   
Hunter College, City 
    University of New York, 695 Park Avenue, New York, NY 10021, 
    USA} 
 
\date{\today} 
  
\begin{abstract}  
  
We consider two different optimized measurement strategies for the discrimination 
of  nonorthogonal quantum states. The first is ambiguous discrimination with a  
minimum probability of inferring an erroneous result, and the second is  
unambiguous, i. e. error-free, discrimination with a minimum probability of getting 
an inconclusive outcome, where the measurement fails to give a definite answer.     
For distinguishing between two mixed quantum states, we investigate the relation  
between the minimum error probability achievable in ambiguous discrimination, and 
the minimum failure probability that can be reached in unambiguous discrimination  
of the same two states. The latter turns out to be at least twice as large as the  
former for any two given states. As an example, we treat the case that the state of  
the quantum system is known to be, with arbitrary prior probability, either a given  
pure state, or a uniform statistical mixture of any number of mutually orthogonal  
states. For this case we derive an analytical result for the minimum probability of  
error and perform a quantitative comparison to the minimum failure probability. 
\end{abstract}  
 
\pacs{03.67.-a, 03.65.Ta, 42.50.-p} 
\maketitle

\section{Introduction}  
  
Stimulated by the rapid developments in quantum communication  
and quantum cryptography,  
the question as to how to optimally discriminate  between different   
quantum states has gained renewed interest  
\cite{chefles}.  The problem is to determine the  
actual state of a quantum system that is prepared,   
with given prior probability, in a certain but unknown state   
belonging to a finite set of given possible states.   
When the possible states are not mutually orthogonal, it is impossible to   
devise a measurement that can distinguish between them perfectly.   
Therefore optimum measurement strategies have been developed with respect   
to various criteria. 
  
Recently much work has been devoted to the strategy  
of optimum unambiguous discrimination.   
Here it is required that, whenever a definite outcome is returned after   
the state-distinguishing measurement, the result should   
be error-free, i. e. unambiguous. This can be achieved at the expense of allowing    
for a non-zero probability of inconclusive outcomes, where the measurement fails   
to give a definite answer. When the probability of failure is minimum, optimum unambiguous  
discrimination is realized.     
Analytical solutions for the minimum failure probability, $Q_F$, have been found for  
distinguishing between   
two \cite{ivanovic,dieks,peres,jaeger}  and three \cite{duan,terno,ysun1}   
arbitrary pure states,  
and between any number of pure states that are symmetric and equiprobable    
\cite{chefles2}.    
On the other hand, the investigation of unambiguous discrimination   
involving mixed states, or sets of pure states, respectively,   
started only recently \cite{ZY,SBH,BHH,jex,rudolph,raynal,eldar4}.    
So far exact analytical results are known only for simple cases  
\cite{SBH,BHH,jex,rudolph}.   
In addition, for unambiguously discriminating between two arbitrary   
mixed states, general upper and lower bounds have 
been derived for the minimum failure probability \cite{rudolph}.    
  
In contrast to unambiguous discrimination, the earliest measurement   
strategy for distinguishing nonorthogonal quantum states    
requires that a definite, i. e. conclusive outcome   
is to be returned in each single measurement. This means that errors   
in the conclusive result are unavoidable and the discrimination is ambiguous.    
Based on the outcome of the measurement, a guess is made as  
to what the state of the quantum system was.   
The optimum measurement then minimizes the probability of  
errors,  i. e. the probability of making a wrong guess.  
For distinguishing two mixed quantum states,   
a general expression for the minimum achievable error   
probability, $P_E$, has been derived in the pioneering work     
by Helstrom \cite{helstrom}.  
When more than two given states are involved,  
an analytical solution is known only for a restricted number of cases   
 the most important  
of them being the case of equiprobable and symmetric  
states that are either pure \cite{ban} or mixed \cite{eldar1,chou}.  
Finally it is worth mentioning that the   
original minimum-error discrimination strategy has been extended   
to determine the minimum achievable probability  
of errors under the condition that a  
fixed finite probability of inconclusive outcomes  
is allowed to occur \cite{barnett,fiurasek}, giving no definite result.  
 
In the present contribution we investigate the relation between the   
minimum error probability, $P_E$, for ambiguously distinguishing two   
mixed quantum states,  
and the minimum failure probability, $Q_F$,    
attainable in unambiguous discrimination of the same two states. 
In Section II we show that for two arbitrary mixed quantum states   
the latter is always at least twice as large as the  
former. As an analytically solvable special example,   
in Section III we treat the problem of deciding whether   
the state   
of the quantum system is either a given pure state,   
or a  
 mixed state being a uniform statistical mixture of   
any number of mutually orthogonal states.   
First we derive an analytical expression for the minimum error   
probability in this example, extending a previous result   
\cite{H} to the case of arbitrary prior probabilities.   
We then perform the comparison with unambiguous discrimination by  
making use of the general solution for the minimum failure probability   
in unambiguous quantum state filtering out of an arbitrary number   
of states \cite{BHH}. Note that in our preceding work we considered    
state discrimination involving mixed states in the context of    
distinguishing between two sets of pure states,  
referring to the discrimination problem as    
filtering \cite{HB,SBH,BHH}  
when the first set contains only a single state.    
Apart from being an illustration for the general    
relation between $P_E$ and $Q_F$, our specific example is of interest on its own   
for applications that are mentioned in the conclusions.

\section{Inequality for the minimum probabilities   
of error and of failure}   
 
In the frame of the quantum detection and estimation    
theory \cite{helstrom}, a measurement that discriminates between two   
mixed states, described by the density operators   
$\rho_1$ and $\rho_2$, and occurring with the prior    
probabilities $\eta_1$ and $\eta_2=1-\eta_1$, respectively,  
can be formally      
described with the help of two detection operators    
${\Pi}_1$ and  ${\Pi}_2$.  These operators are defined in such a   
way that    
${\rm Tr}({\rho}{\Pi}_j)$ is the probability to infer the system   
is in the state $\rho_j$ if it has been prepared in the state   
$\rho$.   
Since the probability is a real non-negative number, the   
detection operators have to be Hermitean and positive-semidefinite.   
In the error-minimizing measurement scheme    
the measurement is required to be exhaustive and conclusive in the sense  
that in each single case with certainty one of the two possible states is    
identified, although perhaps incorrectly, while inconclusive results 
allowing no identification do not occur. This leads to the requirement 
\begin{equation}   
 {\Pi_1} + {\Pi_2} = I_{D_S},   
\label{-1}   
\end{equation}   
where $I_{D_{S}}$ denotes the unit operator     
in the $D_S$-dimensional  physical state space of the quantum system     
under consideration.  
The overall probability $P_{\rm   
  err}$ to make an erroneous guess for any of the incoming states is    
then given by   
\begin{equation}   
P_{\rm err}=1-\sum_{j=1}^2\eta_j{\rm Tr}({\rho}_j{\Pi}_j) =    
\eta_1{\rm Tr}({\rho}_1{\Pi}_2) +    
\eta_2{\rm Tr}({\rho}_2{\Pi}_1),   
\label{-2}   
\end{equation}   
where use has been made of the relation $\eta_1 + \eta_2 = 1$.  
In order to find the strategy for    
minimum-error discrimination, one has to determine the specific    
set of detection operators that minimizes the value of $P_{\rm err}$     
under the constraint given by Eq. (\ref{-1}).    
As found by Helstrom \cite{helstrom},     
the smallest achievable error probability   
$P_{\rm err}^{\rm min}=P_E$ is given by   
\begin{equation}   
P_E = \frac{1}{2}\left(1 - {\rm Tr} |\eta_2 \rho_2 - \eta_1   
{\rho}_1|\right),    
\label{-hel}   
\end{equation}   
where $|\sigma| =   
\sqrt{\sigma^{\dag}\sigma}$ for any operator $\sigma$.    
  
While the original derivation of Eq. (\ref{-hel}) relies on variational   
techniques,    
for the purpose of this paper it is advantegeous to analyze the   
two-state minimum-error measurement with the  
help of an alternative method \cite{fuchs,virmani}.  
To this end we express Eq. (\ref{-2}) alternatively as   
\begin{equation}  
P_{\rm err}= \eta_1 + {\rm Tr}({\Lambda}{\Pi}_1)  
= \eta_2 - {\rm Tr}({\Lambda}{\Pi}_2),   
\end{equation}  
where we introduced the Hermitean operator   
\begin{equation}  
\Lambda= \eta_2 {\rho}_2 - \eta_1 {\rho}_1  
= \sum_{k=1}^{D_S} \lambda_k |\phi_k\rangle \langle \phi_k|.  
\label{-lambda}  
\end{equation}  
Here the states $|\phi_k\rangle$ denote the orthonormal eigenstates  
belonging to the eigenvalues $\lambda_k$ of the operator $\Lambda$. By  
using the spectral decomposition of $\Lambda$, we get the representations   
\cite{H}  
\begin{equation}  
P_{\rm err}=\eta_1 +   
\sum_{k=1}^{D_S} \lambda_k \langle \phi_k |{\Pi}_1|\phi_k\rangle   
=\eta_2 - \sum_{k=1}^{D_S} \lambda_k \langle \phi_k  
|{\Pi}_2|\phi_k\rangle.   
\label{-P}  
\end{equation}  
The eigenvalues $\lambda_k$ are real, and without loss of generality  
we can number them in such a way that   
\begin{eqnarray}  
\lambda_k   <  0 \qquad & {\rm for}& \qquad 1 \leq k < k_0 \nonumber\\   
\lambda_k   >  0 \qquad & {\rm for} & \qquad  k_0 \leq k \leq D  
\nonumber\\   
\lambda_k = 0 \qquad  & {\rm for} & \qquad D < k \leq D_S.   
\end{eqnarray}  
The optimization task is then to determine the specific operators  
${\Pi}_1$, or ${\Pi}_2$, respectively, that minimize the right-hand  
side of Eq. (\ref{-P}) under the constraint that   
\begin{equation}  
0\leq \langle \phi_k |{\Pi}_j|\phi_k\rangle \leq 1  
\qquad (j=1,2)   
\end{equation}  
for all eigenstates $|\phi_k\rangle$. The latter requirement is due to  
the fact that ${\rm Tr}({\rho}{\Pi}_j)$ denotes a probability for any  
${\rho}$. From this constraint and from Eq. (\ref{-P}) it immediately   
follows that the smallest possible error probability, $P_{\rm  
  err}^{\rm min} \equiv P_E$, is achieved when the detection operators  
are chosen in such a way that the equations $\langle \phi_k  
|{\Pi}_1|\phi_k\rangle = 1$ and $\langle \phi_k |{\Pi}_2|\phi_k\rangle  
= 0$ are fulfilled for eigenstates belonging to negative eigenvalues,  
while eigenstates corresponding to positive eigenvalues obey the  
equations $\langle \phi_k |{\Pi}_1|\phi_k\rangle = 0$ and $\langle  
\phi_k |{\Pi}_2|\phi_k\rangle = 1$. Hence the optimum detection  
operators are given by   
\begin{equation}  
{\Pi}_{1}   =  \sum_{k=1}^{k_0 -1}  
|\phi_k\rangle \langle \phi_k|,  
\qquad  
{\Pi}_{2}   =  \sum_{k=k_0}^{D}   
|\phi_k\rangle \langle \phi_k|,   
\label{-op}  
\end{equation}  
where these expressions have to be supplemented by projection  
operators onto eigenstates belonging to the eigenvalue $\lambda_k =  
0$, in such a way that ${\Pi}_{1} + {\Pi}_{2} = {I}_{D_S}$.   
Using Eq. (\ref{-2}),   
from the optimum detection operators the minimum error   
probability is found to be \cite{H} 
\begin{equation}  
P_E = \eta_1 - \sum_{k=1}^{{k_0}-1}|\lambda_k|   
= \eta_2 - \sum_{k=k_0}^D|\lambda_k|.  
\end{equation}  
By taking the sum of these two alternative representations, using  
$\eta_1 + \eta_2 = 1$, we arrive at   
\begin{equation}  
P_E = \frac{1}{2}\left(1 - \sum_{k}|\lambda_k|\right)   
= \frac{1}{2}\left(1 - {\rm Tr}|\Lambda|\right)   
\label{-hel1}  
\end{equation}  
which is equivalent to Eq. (\ref{-hel}).  
Interestingly, for characterizing the measurement    
described by the detection operators given in Eq. (\ref{-op}),   
two different cases have to be considered.   
Provided that there  
are positive as well as negative eigenvalues in the spectral  
decomposition of ${\Lambda}$, the measurement obviously   
is a von Neumann measurement that  
consists in performing projections onto the two orthogonal subspaces  
spanned by the two sets of states   
$\{|\phi_1\rangle,\dots,|\phi_{k_0-1}\rangle \}$  
and $\{|\phi_{k_0}\rangle,\dots,|\phi_{D}\rangle \}$.  
On the other hand,   
when negative eigenvalues do not exist it follows that  
${\Pi}_1= 0$ and ${\Pi}_2={I}_{D_S}$ which means that the minimum  
error probability can be achieved by always guessing that the quantum  
system is in the state ${\rho}_2$, without performing any  
measurement at all. Similar considerations hold true in the absence of  
positive eigenvalues. These findings are in agreement with the  
recent observation \cite{hunter} that a measurement does not  
always aid minimum-error discrimination. In Section III   
we shall discuss a corresponding example.   
  
In the error-minimizing scheme for discriminating  two mixed  
states $\rho_1$ and $\rho_2$ of a quantum system, a non-zero  
probability of making a correct guess can always be achieved. However,  
it is obvious that the states can only be distinguished   
unambiguously when at least one of the mixed states contains at  
least one component, in the $D_S$-dimensional physical state space of  
the quantum system, that does not also occur in the other mixed  
state. As has been shown recently \cite{rudolph}, the minimum failure   
probability in unambiguous   
discrimination, $Q_F$, obeys the inequality      
\begin{equation}  
Q_F \geq   \left\{ \begin{array}{ll}  
      2\sqrt{\eta_1 \eta_2}\,F(\rho_1,\rho_2) &  
\qquad  
        \mbox{if $  
       \sqrt{\frac {\eta_{\rm min}}{\eta_{\rm max}}} \geq F$} \\  
      \eta_{\rm min} + \eta_{\rm max} [F(\rho_1,\rho_2)]^2   
        &   
\qquad  
\mbox{otherwise}.       
                  \end{array}  
     \right.  
\label{Q}   
\end {equation}  
Here $\eta_{\rm min}$ $(\eta_{\rm max})$ is the smaller (larger) of the two   
prior probabilities $\eta_1$ and $\eta_2$, and $F$ is the fidelity,   
defined as   
\begin{equation} 
F({{\rho}_1, {\rho}_2})= {\rm Tr}[   
\left(\sqrt{{\rho}_2}\; {\rho}_1 \sqrt{{\rho}_2}\right)^{1/2}].  
\label{f} 
\end{equation} 
Since the two lines of Eq. (\ref{Q}) are the geometric and the arithmetic  
mean, respectively, of the same expressions, it is clear that    
the first line denotes the overall lower bound, $Q_L$,  
on the failure probability, i. e.   
\begin{equation}  
Q_F \geq Q_L \equiv 2\sqrt{\eta_1 \eta_2}\,F({{\rho}_1, {\rho}_2})   
\label{f1}  
\end{equation}  
for arbitrary values of the prior probabilities.  
 
In the following we want to compare the minimum error probability,   
$P_E$, given by Eq. (\ref{-hel}) with the smallest possible failure   
probability that is   
achievable in a measurement designed for discriminating the two mixed   
states unambiguously.  
Our procedure will be closely related to the   
derivation of inequalities   
between the fidelity and the trace distance \cite{nielsen}.  
In order to estimate  
$Q_L$, or the fidelity, respectively, it is advantageous  
to use a particular orthonormal basis. It has been proven   
\cite{fuchs,nielsen} that when the  
basis states are chosen to be the eigenstates $\{|l\rangle\}$ of the  
Hermitean operator ${\rho}_2^{-1/2}  
\left(\sqrt{{\rho}_2}\; {\rho}_1 \sqrt{{\rho}_2}\right)^{1/2}   
{\rho}_2^{-1/2}$, the fidelity takes the  
form   
\begin{equation}  
F({{\rho}_1, {\rho}_2}) =   
\sum_{l}\sqrt{\langle l |{\rho}_1|l\rangle   
\langle l |{\rho}_2|l\rangle} = \sum_l \sqrt{r_l \;s_l}.   
\end{equation}  
Here $\sum{_l} |l\rangle\langle l|= {I}$,   
with $I$ being the unit operator, and we introduced the  
abbreviations $r_l = \langle l |{\rho}_1|l\rangle$ and $s_l = \langle  
l |{\rho}_2|l\rangle$. The lower bound on the failure probability then  
obeys the equation  
\begin{equation}  
1-Q_L =1-  2 \sqrt{\eta_1 \eta_2}\;\sum_l \sqrt{r_l\;s_l}   
= \sum_l \left(\sqrt{\eta_1 r_l}-\sqrt{\eta_2 s_l}\right)^2,  
\label{-QL}  
\end{equation}  
where the second equality sign is due to the relation $\eta_1 + \eta_2  
=1$ and to the normalization conditions ${\rm Tr} \rho_1 = \sum_l r_l  
= 1$ and ${\rm Tr} \rho_2 = \sum_l s_l = 1$.  
  
Let us now estimate the minimum error probability $P_E$, using the   
same set of basis states $\{|l\rangle\}$. Because of Eq. (\ref{-hel1}) and   
of the fact that $ \langle \phi_k |\phi_k \rangle =  \sum_l |\langle  
\phi_k |l\rangle|^2 =1$, we can write  
\begin{eqnarray}  
1 - 2 P_E  & = & \sum_{k} |\lambda_k| =   
\sum_{l} \sum_k|\lambda_k| |\langle \phi_k |l\rangle|^2  \nonumber\\   
 &  \geq  & \sum_l \left|\sum_k\lambda_k  |  
             \langle \phi_k |l\rangle|^2 \right|   
= \sum_l |\langle l |{\Lambda}|l \rangle|,   
\end{eqnarray}  
where the last equality sign follows from the spectral decomposition  
of the operator ${\Lambda}$, see Eq. (\ref{-lambda}).   
After reexpressing ${\Lambda}$ in  
terms of the density  
operators describing the given states, we arrive at  
\begin{eqnarray}  
1-2 P_E  & \geq  & \sum_{l} \left|\langle l|\eta_1 \rho_1   
 - \eta_2 \rho_2 |l \rangle\right| \nonumber\\   
& = & \sum_l   
\left|\sqrt{\eta_1 r_l}-\sqrt{\eta_2 s_l}\right|   
\left|\sqrt{\eta_1 r_l}+\sqrt{\eta_2 s_l}\right|.   
\label{-PE}  
\end{eqnarray}  
By comparing the expressions on the right-hand sides of  
Eqs. (\ref{-QL}) and (\ref{-PE}) it becomes  
immediately obvious that $1- 2 P_E \geq 1 - Q_L$, 
or $P_E \leq Q_L/2$, respectively.   
Together with Eq. (\ref{f1}) this implies our final result    
\begin{equation}  
P_E \leq  \frac{1}{2} Q_F.  
\label{ineq}  
\end{equation}  
Hence for two arbitrary mixed states, occurring with arbitrary prior   
probabilities,  the smallest possible failure  
probability in unambiguous discrimination is at least twice as large  
as the minimum probability of errors achievable for ambiguously  
distinguishing the same states.

\section{Distinguishing between a pure state and a uniformly mixed state}  
    
For a quantitative comparison between the minimum probabilities of   
error and of failure we wish to consider a state discrimination problem   
that involves mixed states and that can be solved   
analytically with respect to the two different strategies 
under investigation.   
Minimum-error discrimination between two mixed states, or   
between two sets of states consisting both of a certain number    
of given pure states, respectively, has been recently treated analytically      
under the restriction that the total Hilbert space    
collectively spanned by the states is only two-dimensional \cite{HB}.   
When the dimensionality $D$ of the relevant Hilbert    
space is larger than two, however,   
the explicit analytical evaluation of $P_E$  
poses severe difficulites, due to the fact that applying the    
Helstrom formula amounts to calculating the eigenvalues of a    
$D$-dimensional matrix.     
In the following we consider a simple yet non trivial    
discrimination problem where we are able to find an analytical solution   
for minimum-error discrimination    
in an Hilbert space of arbitrary many dimensions.  
  
We assume that the quantum system is either prepared, with the  
prior probability $\eta_1$, in the pure state  
\begin{equation} 
\rho_1 = |\psi\rangle \langle \psi|,  
\label{10} 
\end{equation} 
or, with the prior probability   
$\eta_2 = 1 - \eta_1$, in a uniform statistical  
mixture of $d$ mutually orthonormal states,   
described by the density operator  
\begin{equation} 
\rho_2= \frac{1}{d}\sum_{j=1}^{d}|u_j\rangle \langle u_j| 
=\frac{1}{d}\;I_d 
\label{11}  
\end{equation} 
with $\langle u_i|u_j\rangle = \delta_{ij}$ and 
$I_d$ denoting the unit operator in the $d$-dimensional Hilbert space  
$\mathcal{H}_d$ spanned by the states  
$|u_1\rangle,\ldots,|u_{d}\rangle$. 
It is convenient to introduce additional  
mutually orthogonal and normalized states  
$|v_0\rangle$ and $|v_1\rangle$  
in such a way that 
\begin{equation} 
|\psi\rangle =  
\sqrt{1-\|\psi^{\parallel}\|^2}\;|v_0\rangle\;   
+ \; \|\psi^{\parallel}\|\;|v_1\rangle, 
\label{12} 
\end{equation} 
where $\|\psi^{\parallel}\|\;|v_1\rangle \equiv|\psi^{\parallel}\rangle$  
is the component of  
$|\psi\rangle$ that lies in $\mathcal{H}_d$, i. e.  
\begin{equation} 
\|\psi^{\parallel}\|^2 =  
\langle \psi^{\parallel}|\psi^{\parallel}\rangle = 
\sum_{j=1}^{d}|\langle u_j|\psi\rangle|^2.   
\label{13} 
\end{equation} 
The total Hilbert space spanned by the set of   
states $\{|\psi\rangle,|u_1\rangle,\ldots, 
|u_{d}\rangle\}$ is $d$-dimensional if $\|\psi^{\parallel}\|=1$, and 
$(d+1)$-dimensional otherwise. 
 With $D_S$ denoting the  
dimensionality  of the physical state space of the quantum system 
under consideration, it is clear that the  
relations $D_S \geq d$ or $D_S \geq d+1$ have to be fulfilled in the 
former and the latter case, respectively.   
 
In order to calculate the minimum error probability $P_E$  
with the help of Eq. (\ref{-hel1}),   
we have to  
determine the eigenvalues $\lambda$ of the operator  
$\Lambda= \eta_2 \rho_2 - \eta_1\rho_1$. This amounts to solving the  
characteristic equation ${\rm det}A=0$ with  
\begin{equation} 
A(\lambda)=\lambda I_{d+1} -\Lambda= 
\lambda I_{d+1} + \eta_1|\psi\rangle \langle \psi| 
- \frac{\eta_2}{d}\;I_d,  
\label{A} 
\end{equation} 
where the unit operator in $\mathcal{H}_{d+1}$ can be written as  
$I_{d+1} = |v_0\rangle\langle v_0|+I_{d} $.  
We now take advantage of the fact that by changing  
the basis system the unit  
operator in $\mathcal{H}_d$ can be alternatively   
expressed as $I_{d}= |v_1\rangle\langle v_1|  
+\sum_{j=2}^{d} |v_j\rangle\langle v_j|$ 
with $|v_1\rangle$ being given by Eq. (\ref{12}) and  
$\langle v_i|v_j\rangle = \delta_{ij}$ for $i,j= 0,1,\ldots, d$. 
Therefore 
\begin{equation} 
A = \lambda |v_0\rangle\langle v_0| + \eta_1|\psi\rangle \langle \psi| 
+ \left( \lambda - \frac{\eta_2}{d}\right) 
\sum_{j=1}^{d} |v_j\rangle\langle v_j|,  
\label{A1} 
\end{equation} 
and by using the decomposition of $|\psi\rangle$ in this basis,  
Eq.\ (\ref{12}), we readily obtain the  
matrix elements $A_{ij} = \langle v_i| A(\lambda)|v_j\rangle$.  
>From the condition  ${\rm det}A=0$ the eigenvalues are found to be      
\begin{eqnarray}  
\lambda_{1,2}  & = &   \frac{1}{2} \left[\frac{\eta_2}{d} - \eta_1  
 \mp  \sqrt{\left(\frac{\eta_2}{d}+ \eta_1\right)^2    
 - 4 \eta_1 \frac{\eta_2}{d} \|\psi^{\parallel}\|^2}\;\right],   
\nonumber\\ 
\lambda_k & = & \frac{\eta_2}{d} \;\;\;\;\;\; (k=3,\ldots d+1). 
\label{21}  
\end{eqnarray}  
Clearly, when the set of   
states $\{|\psi\rangle,|u_1\rangle,\ldots, 
|u_{d}\rangle\}$ is linearly independent,   
i. e. for $\|\psi^{\parallel}\| < 1$,   
the square root in Eq. (\ref{21})   
is larger than $|\eta_2/d -\eta_1|$. Therefore    
$\lambda_1$ is the only negative eigenvalue   
and, according to Eq. (\ref{-op}),  the detection operator $\Pi_1$  
that determines the minimum-error measurement scheme is the projector   
onto the eigenstate belonging to the negative eigenvalue.   
Provided that $\eta_1 > \eta_2/d$, the same holds true   
for $\|\psi^{\parallel}\| = 1$.  
However, when $\|\psi^{\parallel}\| = 1$   
and $\eta_1 \leq \eta_2/d$,   
a negative eigenvalue does not exist,    
implying that $\Pi_1=0$. This means that in the   
latter case one cannot find   
a measurement strategy that yields a smaller probability of    
errors than always guessing that the quantum system is    
prepared in the state $\rho_2$.      
By inserting the eigenvalues (\ref{21}) into Eq. (\ref{-hel1})   
we finally arrive at the minimum error probability   
\begin{equation}    
P_E = \frac{\eta_1}{2}+\frac{\eta_2}{2d} -    
 \frac{1}{2} \sqrt{\left(\eta_1+\frac{\eta_2}{d}\right)^2    
 - 4 \eta_1 \frac{\eta_2}{d} \|\psi^{\parallel}\|^2} \ .   
\label{P}    
\end{equation}    
For $d=1$, Eq. (\ref{P}) reduces to the well-known   
Helstrom bound \cite{helstrom} for minimum-error   
discrimination between the two pure states $|\psi\rangle$ and    
$|u_1\rangle$.     
  
Now we turn to unambiguous discrimination.  
The task of distinguishing without errors between the two states   
given by Eqs. (\ref{10}) and (\ref{11}),  
at the expense of allowing inconclusive results to occur, where the  
procedure fails, is a special case of unambiguous quantum state   
filtering \cite{BHH}.  In the latter problem we want to discriminate  
without error a quantum state $|\psi\rangle$, occurring with the prior  
probability $\eta_1$, from a set of states $\{|\psi_j\rangle\}$,  
with prior probabilities $\eta_j^{\prime}$.     
With the substitution $\eta_j^{\prime} = \eta_2/d$ $(j=1,\ldots, d)$,    
the solution given in \cite{BHH} yields for our  
specific example the minimum failure probability   
\begin{eqnarray}  
    \label{27}  
    Q_F = \left\{   
\begin{array}{ll}  
    2 \sqrt{\eta_1 \frac{\eta_2}{d}}\; \| \psi^{\parallel}\|  
    & \mbox{if  $\| \psi^{\parallel} \|^{4}  
    <  \frac{\eta_2}{\eta_1 d} \;\| \psi^{\parallel} \|^{2}  
     < 1$},   
\vspace{0.2cm} \\     
\eta_{1}+ \frac{\eta_2}{d} \;\| \psi^{\parallel} \|^{2}  
     & \mbox{ if $\frac{\eta_2}{\eta_1 d}\; \| \psi^{\parallel} \|^{2} \geq 1$} ,   
\vspace{0.2cm} \\ 
    \eta_{1}\| \psi^{\parallel} \|^{2}+\frac{\eta_2}{d}  
     & \mbox{ if $\frac{\eta_2}{\eta_1 d} \leq    
         \|\psi^{\parallel} \|^{2}$} ,  
    \end{array}  
    \right.  
\end{eqnarray}  
where $\|\psi^{\parallel} \|$ is defined by   
Eq. (\ref{13}), and where again $\eta_1 + \eta_2 = 1$.  
As shown in \cite{BHH}, the second and the third line  
of Eq. (\ref{27}) refer to two different types of von Neumann measurements     
while the failure probability given in the first line   
can be reached only by a generalized measurement.   
In the following we compare the minimum error probability, $P_E$,   
with the minimum failure probability, $Q_F$, considering several   
special cases.   
 
\begin{figure}[ht]   
\epsfig{file=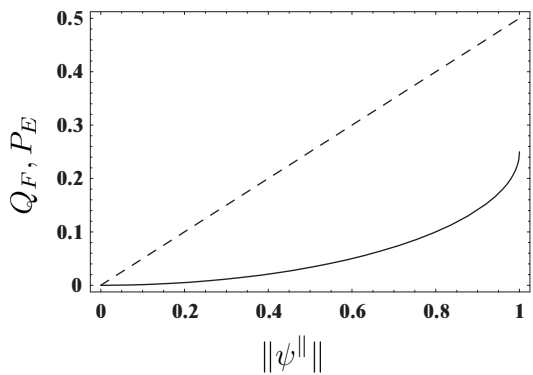,height=5cm,width=7cm}   
\caption{Minimum probabilities of error in ambiguous   
discrimination, $P_E$, (full line), and of    
failure in unambiguous discrimination, $Q_F$, (dashed line),    
for distinguishing between $\rho_1=|\psi\rangle\langle\psi|$   
and $\rho_2 = \frac{1}{3}\sum_{j=1}^3 |u_j\rangle\langle u_j|$,   
where $\langle u_i|u_j\rangle = \delta_{ij}$.   
The probabilities are plotted versus the norm of the parallel component,    
$\|\psi^{\parallel} \| =    
\left(\sum_{j=1}^3|\langle u_j|\psi\rangle|^2\right)^{1/2}$,  
and the prior probability of $\rho_1$ is assumed to be $\eta_1= 0.25$.}   
\label{Fig1}   
\end{figure}   
  
A considerable simplification arises when the states    
$|\psi\rangle,|u_1\rangle,\ldots,   
|u_{d}\rangle$ all occur with the same prior probabilities, i. e. when   
$\eta_1 = \eta_2/d=1/(d+1)$ \cite{H}.   
In this case  Eq. (\ref{P}) yields    
\begin{equation}   
P_E = \frac{1}{d+1}\left( 1 -  \sqrt{1   
- \|\psi^{\parallel}\|^2}\right),   
\label{23}   
\end{equation}   
and from (\ref{27}) we obtain   
\begin{equation}  
Q_F = \frac{2}{d+1}\|\psi^{\parallel} \|,  
\end{equation}  
where the latter expression is valid in the whole range of the  
possible values of $\|\psi^{\parallel} \|$.   
When $\rho_1$ and $\rho_2$ are nearly orthogonal, i. e. when   
$\|\psi^{\parallel} \| \ll 1$, the minimum error probability, $P_E$,   
takes the approximate value $\|\psi^{\parallel} \|^2/(2d+2)$   
and is therefore significantly smaller than the  
minimum failure probability, $Q_F$, that is 
achievable in unambiguous discrimination (see Fig. 1). On the other   
hand, when $\|\psi^{\parallel} \| = 1$ the ratio $Q_F/P_E = 2$ is reached. 
This is an example from which it becomes obvious that the bound given  
by the general inequality (\ref{ineq}) is tight. 
\begin{figure}[ht]   
\epsfig{file=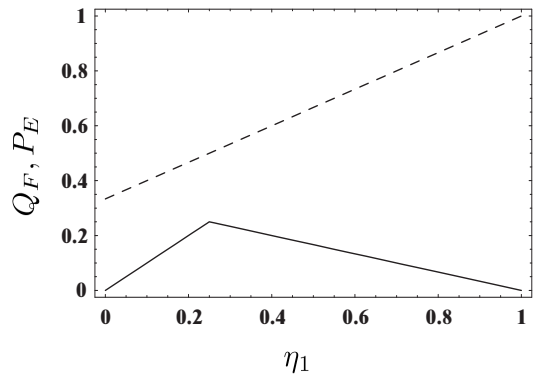,height=5cm,width=7cm}   
\caption{Minimum error probability in ambiguous discrimination,  
$P_E$ (full line), and minimum failure probability in unambigous  
discrimination, $Q_F$ (dashed line), for the states $\rho_1$ and $\rho_2$  
specified in Fig. 1. The probabilities are depicted versus the prior probability    
$\eta_1$ of the state $\rho_1$, in the special case  
that $\|\psi^{\parallel}\| = 1$. 
}    
\label{Fig2}   
\end{figure}   
  
For arbitrary prior probabilities we   
first investigate the discrimination in the linearly dependent case   
$\|\psi^{\parallel} \| = 1$ (see Fig. 2).  
>From Eq. (\ref{P}) with $\eta_2 = 1 - \eta_1$ we then find    
the minimum error probability      
\begin{equation}   
P_E = \frac{1}{2d}\left[ 1 + \eta_1(d-1)    
- |1-\eta_1(d+1)|\right].   
\end{equation}   
Hence as long as $\eta_1 \leq 1/(d+1)$, which is equivalent to  
$\eta_1 \leq \eta_2/d$, we get $P_E = \eta_1$.   
As discussed in connection with the eigenvalues given in   
Eq. (\ref{21}), the best discrimination strategy is then to  
always guess the   
quantum system to be in the state $\rho_2$, and it is not necessary to   
perform any measurement at all.    
However, for $\eta_1 \geq 1/(d+1)$ and  $\|\psi^{\parallel} \| = 1$  
we obtain the minimum error probability   
$P_E = (1-\eta_1)/d$. Now the optimum strategy for minimum-error   
discrimination is to infer the system to be in the state $\rho_1$ when  
the detector along $|\psi\rangle$ clicks which is just  
the eigenstate belonging to the negative eigenvalue, and that the state  
is  $\rho_2$ for a click in any projection onto a direction orthogonal  
to $|\psi\rangle$.   
With $\|\psi^{\parallel} \| = 1$, from Eq. (\ref{27}) the minimum   
failure probability  in unambiguous discrimination follows to be   
\begin{equation}  
Q_F= \eta_1 + \frac{1}{d}(1-\eta_1).  
\label{28}  
\end{equation}  
The strategy for optimum unambiguous discrimination in this case  
is also the von Neumann measurement   
consisting of projections onto the state   
$|\psi\rangle$ and onto the subspace orthogonal      
to $|\psi\rangle$.   
When a click occurs from   
projection onto the orthogonal subspace,   
the state $\rho_2$ is uniquely identified.   
The measurement    
fails to give a conclusive answer when either the state   
$|\psi\rangle $ was present, which occurs  
with probability $\eta_1$, or when the state    
$\rho_2$ was present and a click resulted from   
projection onto $|\psi\rangle$, which occurs with   
the probability $\eta_2/d$.  
 
\begin{figure}[ht]   
\epsfig{file=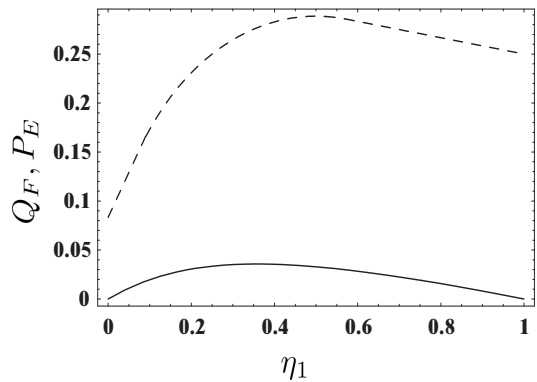,height=5cm,width=7cm}   
\caption{Same as Fig. 2 but for the case $\|\psi^{\parallel} \| = 0.5$.}    
\label{Fig3}   
\end{figure}   
  
Finally in Fig. 3 an example is depicted   
for arbitrary prior probabilites and     
linearly independent states, where $\|\psi^{\parallel} \| < 1$.   
Obviously the minimum error probability, $P_E$, given by    
Eq. (\ref{P}) is in general much smaller than    
the minimum failure probability, $Q_F/2$, given by Eq. (\ref{27}).

\section{Conclusions} 
 
We showed that the    
minimum error probability, $P_E$, for ambiguously distinguishing any two    
mixed quantum states without inconclusive results, is always at most half as large as    
the minimum failure probability, $Q_F$, for unambiguous, i. e. error-free  
discrimination of the same two states, at the expense of the occurrence of  
inconclusive results where the measurement fails.  
As an example, we gave an exact analytical solution to   
the problem of determining whether the state    
of the quantum system is either a given pure state occurring  
with arbitrary prior probability, or a uniform statistical mixture of    
any number of mutually orthogonal states.    
Uniformly, i. e. completely mixed states have been considered in the context of  
estimating the quality of a source of quantum states, as   
has been recently discussed in connection with single-photon  
sources, introducing the new measure of suitability \cite{hockney}.  
This measure relies on  
identifying all states that would be useful for the specific  
application, finding a set of states spanning the space of  
the useful states, and then defining a target state as a  
complete mixture of those states. 
If the mutually orthogonal states in the uniform statisitical mixture span the  
entire state space of the quantum system, the mixed state    
describes a totally random state, containing no   
information at all. Discriminating between the pure state     
and the mixed state then amounts to deciding whether the system   
has been reliably prepared in the pure state, or whether the preparation   
has failed \cite{hunter}.\\   
{\it Acknowledgements:}  
The research of J. B. was partially supported by the Office of Naval  
Research and by a grant from PSC-CUNY. 

\end{document}